\documentclass{article}

\begin{document}

\begin{center}
{\Large\bf Quark Model Estimates of the Structure of the Meson-$N$-$N^*(1/2^{-})$ Transition Vertices\footnote{Supported
in part by Forschungszentrum FZ J\"ulich (COSY)and by CNPq}}
\end{center}
\begin{center}
M. Dillig \footnote{E-mail: mdillig@theorie3.physik.uni-erlangen.de}    \\
Institute for Theoretical Physics III \footnote{preprint FAU-TP3-06/Nr. 06}, University of Erlangen-N\"urnberg, Erlangen, Germany
\end{center}
\begin{center}
S. S. Rocha, G. F. Marranghello, E.F. L\"utz   and C. A. Z. Vasconcellos
\\ Instituto de F\'{\i}sica, Universidade Federal do Rio Grande do Sul \\ Porto Alegre, Brazil.
\end{center}

\vspace {4.0cm} 

\begin{abstract}\

We address an actual problem of baryon-resonance dominated meson-exchange processes in the low GeV regime, i.e. the
phase and the structure of meson-$NN^*$ transition vertices. Our starting point is a quark-diquark model for
the baryons (obeying approximate covariance; the mesons are kept as elementary objects), together with the relative
phases for the $NN$ vertices, as determined from low energy $NN$ scattering. From the explicit representation of the
$N$ and $N^*$ baryons, we exemplify the derivation of the coupling constants and form factors of the $NN^*$
$(J={1/2}^-)$ transition vertices for pseudo-scalar, scalar and vector mesons.
\end{abstract}

\newpage
{\bf I. Introduction} \\

One of the most actual topics in nuclear physics in the energy region of a few GeV is the investigation of the
excitation and the structure of excited states of the nucleon, i.e., the investigation of $N^*$ baryon
resonances \cite{1}. One experimental tool to study these excitations is the exclusive (near threshold) production of
selective baryon resonances on the nucleon with hadronic or electromagnetic beams. Currently, such experiments are currently 
vigorously pursued with protons at COSY (up to last year also at CELSIUS) \cite{2} and with electrons at MAMI, ELSA, CEBAF, BES  and HERMES
\cite{3}. \\

As a characteristic feature, heavy meson production, even at threshold, involves
 large momentum transfers of typically 1 GeV/c; consequently, such reactions
probe the short range dynamics of the baryon-baryon system. As a consequence, meson-exchange calculations at such
large momentum transfers are in general not dominated by the single pion-exchange: for a quantitative description,
the exchange of all mesons with masses up to typically 1 GeV, i.e., the exchange of the Goldstone bosons $\pi$,
$\eta$, $K$, of the vector mesons $\varrho$, $\omega$ and $K^*$, and of the $2\pi$-dominated scalar mesons $\sigma$
and $\delta$ have to be included explicitly \cite{4}. As the relative phases and the form factors of the
corresponding meson-$NN^*$ vertices are in general unknown, even in the cleanest single-meson production reaction, such as the
near-threshold $\eta$ and $K^+$ production, which are dominated, respectively, by the excitation of the $N^*$(1535)
and $N^*$(1650) resonances \cite{5,6}, the model predictions depend sensitively on the {\it ad hoc} interplay
between the various meson exchanges \cite{7,8,9}. Examples for near-threshold $\eta$-production show a characteristic
uncertainty up to one order of magnitude in exploring the relative phases between the different mesons\cite {7,10}. An
additional, equally drastic uncertainty is the highly unknown behavior of the different meson-baryon transition
vertices if continued far off-shell into the space-like region.\\

Of course, there are prescriptions in the literature, specifying both diagonal and non-diagonal mesonic nucleon
couplings to baryon resonances with arbitrary spins and positive or negative parity, and the relative phases between
different meson-induced transition vertices \cite{11}. For the transition to negative parity, spin 1/2, baryon
resonances, which we investigate in this note, i.e. for $\lambda NN^*$ $(J^\pi={1/2}^-)$ vertices ($\lambda$ denotes
scalar, pseudo-scalar or vector mesons), the standard recipe is to start from the $\lambda NN$ vertices - which are
expected to be well understood from experimental evidence, such as from NN scattering at low energies \cite{12} - and
add for the transition vertices the negative parity operator $i\gamma_5$, yielding a well defined scheme for the
relative mesonic couplings (of course, the strength of various vertices on and off shell remains completely open).
Already at first inspection such a procedure may be ambiguous depending on the explicit order of implementation
relative to the Dirac matrices in the $\lambda NN$ vertices (while, for example, the extension of the pseudo-scalar (PS) $\pi
NN$ vertex to $\pi NN^*$ is unambiguous, however, a sign ambiguity may already arise for the PV vertex from $\gamma_5 \gamma_\mu \, \neq \,  \gamma_\mu \gamma_5$ (the absolute
coupling constants are related by the equivalence theorem for
the baryons on their mass shell \cite{13});
 similarly on the
SU(3) level mesonic vertices involve D and F components, implying different phases for the $\lambda NN^*$
vertices \cite{14}. \\

In this note we would like to investigate these questions, i. e. the relative
 phases of the different meson exchanges and the structure of the meson - 
NN$^*$ transition form factors in a quark-diquark model for the baryons. In the following chapter we derive the corresponding formalism; the main results are then presented and discussed in chapter 3. Finally, in chapter 4 we close with a summary and an outlook
 for improvements. 

\vspace{1.2cm}

{\bf II. Calculational details and results}\\

In this chapter we formulate a quark-diquark representation for the N and
the N$^*$(1535) and elaborate on the phases and the transition form factors
of the $\lambda N N^*$ vertices.\\

{\bf II.1 Quark-diquark representation of the N and N$^*$.}\\

The recent literature shows various attempts towards the relative
phases of meson-baryon-baryon couplings in general; the most extensive
investigations for the coupling to non strange baryon resonances were
obtained from a self-consistent coupled channel approach to hadron and/or $\gamma$-induced meson production on the nucleon \cite{15} or from a detailed analysis
within the non relativistic constituent quark model \cite{16}. However, opposite to the last reference, it seems questionable to fix in a fairly model independent way 
the vertices $\lambda N N^*$, which still involve an 
overall arbitrary phase for the various resonances; only the coupling to a given resonance is well defined with respect to the phases in $\lambda NN$, which 
are fairly well known. 
Thus we focus here on a specific resonance, the $S_{11}(1535)$ and investigate the relative phases of selected $\lambda NN^*$ couplings and the t-channel continuation of the coupling constants (the same conclusions hold for the
 $S_{11}(1650)$ resonance). As we aim to apply
our findings in a next step to meson exchange models at high momentum
transfers up to 1 GeV/c, our emphasis is to estimate form factors for the
off-shell continuation in a (at least approximately) covariant, but still
economical way. For this end, we
compromise on our quark representation of the interacting hadrons:  we represent baryons as quark-diquark
objects \cite{17}, keeping in this work only the scalar-isoscalar component.  For our purpose such an approximation seems well justified: different 
investigations show a strong dominance of the scalar component for the
electromagnetic form factors of the nucleon up to (and beyond) momentum transfers of 1 GeV/c; in the same momentum range axial diquarks renormalize magnetic
form factors, having, however, no significant influence on their momentum spectrum \cite{18}). As the form factors derived show a smooth momentum spectrum
(see the discussion below), minor modifications of axial diquarks are 
readily taken into account in a slight renormalized range parameter for the
quark-diquark wave function. In addition, as a main advantage, Lorentz 
boosts are easily incorporated in a purely scalar diquark picture (the incorporation of Lorentz boosts is still matter of discussion in the literature \cite{19}). The  mesons we still treat as elementary objects, which couple
directly to the q - (qq) system (one may include a phenomenological form factor to simulate their finite extension).
We expect that our treatment of the meson fields is qualitatively acceptable, as we do not aim at absolute
predictions of $\lambda NN^*$ parameters, but develop simple scaling rules for the N$^*$ vertices, by exploring
 their structure relative to the ground-state baryon, i.e., the
nucleon. For the baryons, the quark-diquark representation allows to incorporate approximate covariance beyond
non-relativistic potential models, being able to include Lorentz-quenching in the radial wave function and the
modification of the small component, which enters explicitly into the leading non-relativistic reduction of the
vertices. Explicitly we work within harmonic confinement, yielding for the $N$ and the $({1/2}^{-})N^*$ resonances,
which are assumed to be pure p-shell excitations of a single quark\cite{19},

\begin{equation}
|N({1/2}^+)> \sim \pmatrix{1\cr\frac{\underline{\sigma}\underline{q}}{(1+\lambda)m}}
e^{-\frac{1}{2a^2}(\frac{z^2}{\lambda^2}+\underline{\rho}^2)}|1/2,\mu>_S| 1/2, 1/2>_T | [(10)(01)]00 >_C
\end{equation}
and
\begin{equation}
|N^*({1/2}^-)> \sim \pmatrix{1\cr\frac{\underline{\sigma}\underline{q}}{2m}} \, r e^{-\frac{r^2}{2{a^*}^2}}|[Y_1({\hat
r})1/2] 1/2,\mu^*>_S|\! 1/2, 1/2>_T \!|\! [(10)(01)]00 \! >_C
\end{equation}

Here $r$, $z$ and $\rho=\sqrt{x^2+y^2}$ refer to the $q-(qq)$ relative distance
$\underline{r}=\underline{r}_q-\underline{r}_{qq}$; the p-wave nature of the $N^*$ is reflected by the  $(l,s)=(1,1/2)$ coupling
to $|j,m>=|1/2,\mu^*>$; finally the baryon size parameters are expected to be of the order $a$, $a^*\sim 2/3-3/4$ fm (in eqs. (1,2) for the completely antisymmetric color wave function the standard SU(3) representation in the Elliot notation 
is used \cite{20}). Above we
simulate Lorentz quenching for a baryon with momentum $Q$ along the z-axis
via

\begin{equation}
\lambda(Q)=M/\sqrt{M^2+Q^2}
\end{equation}

\noindent
(with $M$ being the nucleon mass); with respect to practical applications in mind,
i.e., near threshold meson production with the $N^*$ produced practically in
rest in the (by far) dominant post-emission amplitude\cite{7,8,9,10}, we drop here
minor corrections from Lorentz quenching for the resonance and the nucleon in the final state. Clearly more sophisticated
representations for the $N^*(1/2^-)$ as 3q objects found in the literature, they, however, focus either on baryon spectroscopy \cite{21} or
 only on a very selective decay channel (i. e. the $\eta$ channel, \cite{22}).
As in addition, extensions to include $q {\overline q}$ admixtures in baryons as the 
leading component in a systematic Fock expansion, which
 clearly are of significant importance for baryons in the continuum, as well as  a more detailed and systematic baryon-spectroscopy in a
quark-diquark representation are currently missing, we have to defer the
investigation of these aspects to future work .\\

{\bf II.2. Phases of the $\lambda N N^*$ vertices}\\

With these ingredients we formulate our problem explicitly. Starting point are the relativistic
$\lambda NN$ vertex functions with their well defined phase structure for : for $S(0^+)$, $PS(0^-)$ and $V(1^-)$ mesons
they are given explicitly by \cite{12}:\\

- $\lambda NN$: the covariant forms

\begin{eqnarray}
&L_{PS}&=-ig{\bar \psi}\gamma_5\psi\phi_{PS} \equiv -i\frac{f}{m}{\bar\psi}\gamma_5 \slash{\!\!\!q}\psi\phi_{PS} \nonumber
\\ &L_S&=+g{\bar \psi}\psi\phi_S \nonumber \\
&L_V&=-g{\bar\psi}\slash{\!\!\!\epsilon}\psi-i\frac{f}{2M}{\bar\psi}\sigma^{\mu\nu}(q_\mu\epsilon_\nu-q_\nu\epsilon_\mu)\psi,
\end{eqnarray}
\noindent yield in the leading non-relativistic limit

\begin{eqnarray}
&L_{PS}&=-i\left(\frac{f}{m}\right)\underline{\sigma}\underline{q} \nonumber \\
&L_S&=+g \nonumber \\
&L_V&=-g\epsilon_0+\frac{g}{2M}(\underline{\epsilon}\underline{q}-i\underline{\epsilon}(\underline{\sigma}\times \underline{q}))
\end{eqnarray}

\noindent for a nucleon with $\underline{p}$ and $\underline{p}^\prime\sim 0$ (at threshold) in initial and final
state, respectively, and with $\underline{q}\approx\underline{p}$. Above we followed the standard notation from
Bjorken-Drell \cite{23} (we represented the vector fields by the polarization vector $\epsilon_\mu$ with
$\epsilon_\mu\epsilon_\nu=-g_{\mu\nu}$) and keep for the transition to the $N^*$ the leading
$\underline{\epsilon}$-dependence in the vector coupling). From above we formulate with the recipe

\begin{equation}
({\bar\psi}\Gamma\psi)_{NN}\rightarrow i({\bar\psi}\gamma_5\Gamma\psi)_{NN^*}
\end{equation}

\noindent
the vertex functions involving the ($1/2^-$) baryon resonances, yielding for the\\

- $\lambda NN^*$ vertices in the covariant form

\begin{eqnarray}
&L_{PS}^*&=+g^*{\bar\psi}\psi\phi_{PS}=+\left(\frac{f^*}{m}\right)_{PS}{\bar\psi}\slash{\!\!\!q} \psi\phi
\nonumber
\\ &L_S^*&=+ig^*{\bar\psi}\gamma_5\psi\phi_S \nonumber \\ &L_V^*&=-ig^*{\bar\psi}\gamma_5\slash{\!\!\!\epsilon}\psi
\end{eqnarray}

\noindent
with the static limits:

\begin{eqnarray}
&L_{PS}^*&=g^* \nonumber \\ &L_S^*&=+i\left(\frac{f^*}{m}\right)\underline{\sigma}\underline{q} \nonumber \\
&L_V^*&=+i\left(\frac{f^*}{m}\right) \epsilon_0\underline{\sigma}\underline{
q}-ig_V^*\underline{\epsilon}\underline{\sigma}
\end{eqnarray}

\noindent (above we dropped the corresponding tensor term for the $\lambda NN^*$
 vertex, as there is no evidence for a
large anomalous magnetic moment of the $N^*$(1535) or the $N^*$(1650)).\\

For the investigation of the $\lambda NN^*$ vertices in the quark-diquark model we first establish the relative signs of the $\lambda NN$ and
 $\lambda NN^*$ vertices, and supplement the findings in a
second step in a more detailed derivation of the corresponding $N$ and $N^*$ transition form factors.\\

We derive the vertex structure explicitly in momentum space: from the general
structure

\begin{equation}
\underline{F}(\underline{q})=\int \psi^*(\underline{r})\Omega(-i\underline{\nabla})\psi(\underline{r}) e^{i\underline{q}\underline{r}}\underline{dr}
\end{equation}

\noindent
in $r$-space, where $\Omega(\underline{\nabla})$ refers to the operators in eqs.(5,8), we obtain with the corresponding
Fourier transforms

\begin{equation}
\psi(\underline{r})=\frac{1}{(2\pi)^{3/2}}\int e^{i\underline{q}^\prime\underline{r}}\phi(\underline{q}^\prime)\d underline{q}^\prime
\end{equation}

\noindent
together with

\begin{equation}
\frac{1}{(2\pi)^3}\int e^{-i \underline{k} \underline{r}}\Omega(\underline{\nabla})e^{i(\underline{q}^\prime +\underline{q})\underline{r}}\underline{dr}=
\Omega(\underline{q}^\prime +\underline{q})\delta(\underline{q}^\prime +\underline{q}-\underline{k})
\end{equation}

\noindent
and upon evaluating the $\underline{dk}$ integration via the $\delta$-function, the representation

\begin{equation}
\underline{F}(\underline{q})=\int \phi^*(\underline{q}^\prime)\Omega(\underline{q}^\prime)\phi(\underline{q}^\prime-\underline{q})\underline{dq}^\prime
\end{equation}

\noindent
for an arbitrary $\Omega (\underline{q})$.\\

Upon normalizing the momentum space wave functions appropriately we find for the $N$
 with external longitudinal momentum $\underline{Q}=(0,0,Q)$ (as reflected in the coefficient $\lambda(Q)$ from eq. (3)) and the $N^*$
in rest from the coordinate space representations in eqs. (1,2)

\begin{eqnarray}
\phi_N(q,\underline{q}_\perp;Q)=Ne^{-a^2\lambda(Q)^2(q_z-Q)^2/2-a^2\underline{q}^2_\perp/2}|1/2\mu> \nonumber \\
\phi_{N^*}(\underline{q})=N^*e^{-{a^*}^2
q^2/2}q|[Y_1
(\underline{{\hat q}})
1/2]1/2\mu^*>
\end{eqnarray}

\noindent
with

\begin{equation}
N=\sqrt{\lambda(Q)}a^{3/2}/\pi^{3/4}, \hspace{0.2cm}
N^*=-i\sqrt{\frac{8}{3}}\frac{{a^*}^{1/2}}{\pi^{1/4}}
\end{equation}

\noindent
(where, for a practical comparison with $\lambda NN^*$ vertices as obtained by
the $i\gamma_5$ substitution above, we keep the complex unit $(-i)$ in the $N^*$
normalization constant).\\

We establish the relative signs of the $\lambda NN^*$ vertices in the limit of small or vanishing momentum
transfer q; keeping only the relevant invariants, we obtain from

\begin{equation}
M_{\lambda NN^*}(\underline{q})\approx -ic\int
e^{a^2\underline{q} \, \underline{q}^\prime}<[Y_1(\underline{{\hat q}^\prime})1/2]1/2,\mu^*|\Omega(\underline{q}^\prime)|1/2,\mu>d\underline{q}^\prime
\end{equation}

\noindent
with the positive constant c, the following relations to the order of ${\cal O}(Q^0)$
and ${\cal O}(\underline{Q^1})$:\\

i) $\Omega(\underline{q})=1$

\begin{eqnarray}
M(\underline{q})&\sim &-ic\int e^{a^2\underline{q} \, \underline{q}^\prime}<[Y_1(\underline{{\hat q}^\prime})1/2]1/2,\mu^*|1/2,\mu>d\underline{\hat q}^{\prime} \nonumber \\
&\rightarrow &-ic<10, 1/2,\mu|1/2,\mu>qY^2_{10}(\underline{{\hat q}^\prime})<1/2,\mu|1/2,\mu> \nonumber \\
&\rightarrow &-icq(-1)^{1/2+\mu}(-1)^{1/2-\mu}<1/2\mu|\sigma_0|1/2\mu> \nonumber \\
\equiv +ic\underline{\sigma} \underline{q};
\end{eqnarray}

Thus, the scalar NN vertex transforms in leading order as

\begin{equation}
SNN=+g\rightarrow SNN^*=i\frac{f^*}{m}\underline{\sigma}\underline{q}
\end{equation}

ii) $\Omega(\underline{q})=\underline{\sigma}\underline{q}$

\begin{eqnarray}
M(\underline{q})\sim -ic\int <[Y_1(\underline{\hat q}^\prime)1/2]1/2,\mu^*|\underline{\sigma}\underline{\hat q}^\prime|1/2,\mu>d\underline{\hat q}^\prime
\end{eqnarray}

\noindent
(here we may drop even the leading piece the q-dependence) yielding via the
coupled matrix element
\begin{eqnarray}
& &<[Y_1(\underline{{\hat q}})1/2]1/2,\mu^*|[Y_1(\underline{{\hat q}})\underline{\sigma}]^{00}|1/2\mu >= \nonumber \\
&=\, 2 &\pmatrix{1&1/2&1/2\cr 1&1&0 \cr 0&1/2&1/2}[Y_1({\hat
q})Y_1(\underline{{\hat q}})]^{00}<[1/2,\underline{\sigma}]1/2,\mu^*|1/2,\mu >
\end{eqnarray}

\noindent
(we follow the phase conventions of
Edmonds \cite{24}) together with

\begin{equation}
<[1/2,\underline{\sigma}]1/2,\mu^*|1/2\mu>=\delta_{\mu\mu^*}
\end{equation}

\noindent
finally

\begin{equation}
M(\underline{q})=+ic
\end{equation}

\noindent and thus explicitly for pseudo-scalar mesons

\begin{equation}
PSNN=-i\left(\frac{f}{m}\right)\underline{\sigma} \underline{q}\rightarrow PSNN^*=+g^*
\end{equation}

\noindent
to leading order.\\

iii) $\Omega(\underline{q})=\epsilon_0$ and
$\Omega(\underline{q})=\underline{\epsilon} (\underline{\sigma}\times \underline{q})$ for the vector mesons.\\

The coupling procedure follows the examples shown above: For the term proportional to $\epsilon_0$, the quantity
$\underline{\sigma}\underline{q}$ has to be re-introduced for the full operator structure. For the more evolved
piece

\begin{equation}
<[Y_1(\underline{{\hat q}})1/2]1/2,\mu^*|\underline{\epsilon}(\underline{\sigma}\times \underline{q})|1/2,\mu>
\end{equation}

\noindent
(where the leading $\underline{q}$-dependence may be dropped again) we recouple with the identity

\begin{equation}
\underline{\epsilon}(\underline{\sigma}\times\underline{q})=-i\sqrt{6}\, [\underline{\epsilon}[\underline{q}\underline{\sigma}]^{1}]^{00}
\end{equation}

\noindent
the $N^*$ angular momentum to the invariant $\sim \underline{\epsilon} \, \underline{\sigma}$.\\

As the final result we confirm the $i\gamma_5$ - recipe for the magnetic term,

\begin{equation}
VNN=-i\frac{f}{m}\underline{\epsilon} (\underline{\sigma}\times \underline{q})\rightarrow VNN^*=-ig^*\underline{\epsilon} \, \underline{\sigma},
\end{equation}

\noindent
however, we find the opposite sign for the piece proportional to $\epsilon_0$

\begin{equation}
(VNN)_0=-g\epsilon_0\rightarrow VNN^*=-i\left(\frac{f^*}{m}\right)\epsilon_0\underline{\sigma} \underline{q}.
\end{equation}

\noindent (opposite to eq.(8)).
 As a consequence we find different interference pattern of the $\varrho$,
$\omega$ themselves and also relative to the scalar and pseudo-scalar MEC, provided the leading piece of the vector
$\omega NN$ coupling is kept. \\

{\bf II.3. Transition form factors}\\

The explicit evaluation of the baryon form factors
follows similar lines as above for the coupling constants, is, however, more evolved due to the deformation
of the $N$ wave function along the Q-axis in the initial state. Consequently, from the non-spherical
nature the integration over the internal momenta has to be performed separately for longitudinal and
transversal components. Without further detailing the basic formula involves the structure

\begin{eqnarray}
I(q)&=&\int e^{-{a^*}^2{q^\prime}^2/2-a_\lambda^2(q_z^\prime -q)^2-a^2\underline{q}^2_\perp /2}d{q_z^\prime}d\underline{q}_\perp^\prime \nonumber \\
&=&e^{-\frac{a_{\lambda}^2}{2}q^2}\int e^{-\frac{1}{2}(a^2+{a^*}^2)\underline{{q}^\prime}^2_\perp}d\underline
{q}_\perp^\prime \int e^{-\frac{1}{2}({a^*}^2+a^2_\lambda){{q_z}^\prime}^2
+i(-ia_\lambda^2{q_z^\prime})}dq_z^\prime
\end{eqnarray}

\noindent
(with $a_{\lambda}(Q) \equiv a_{\lambda} = \lambda(Q) a$), which can be easily evaluated to yield explicitly

\begin{equation}
I(q)=\frac{\pi^{3/2}}{{b^*}^2b^*_\lambda}e^{-\frac{{b^*}^2}{2}(1- \frac{a^4_\lambda}{2{b^*}^2{b^*_\lambda}^2})q^2}
\end{equation}

\noindent
with

\begin{equation}
{b^*}^2=\frac{1}{2}({a^*}^2+a^2); \, {b^*_\lambda}^2=\frac{1}{2}({a^*}^2+ a^2_\lambda) \, ,
\end{equation}

\noindent
which reduces in the spherical limit for $a=a^*$ to the standard expression

\begin{equation}
I(q)\rightarrow\frac{\pi^{3/2}}{a^3}e^{-\frac{a^2 q^2}{4}}
\end{equation}

Following the recoupling schemes represented above, we represent the vertices from eq.(5) and eq.(8) in terms of 4
different form factors

\begin{eqnarray}
L_{PSNN^*}&&=+g_{PS}^*F_{PS}^*(q^2) \nonumber \\
L_{SNN^*}&&=+i\frac{f_s^*}{m_s}\underline{\sigma} qF_S^*(q^2) \nonumber \\
L_{VNN^*}&&=-i\frac{f_v^*}{m_v}\epsilon_0\underline{\sigma}qF_V^*(q^2) -
ig_v^*\underline{\epsilon}\underline{\sigma}F_T^*(q^2).
\end{eqnarray}

Here the various form factors are linked as follows (form factors without a star
refer to the corresponding NN vertices)

\begin{eqnarray}
F_V(q^2)=F_S(q^2) \nonumber \\
F_T(q^2)=F_{PS}(q^2) \nonumber \\
F_V^*(q^2)=F_S^*(q^2) \nonumber \\
F_T^*(q^2)=F_{PS}^*(q^2).
\end{eqnarray}

With the relevant overlap integrals listed in the appendix we find with the
normalization $F(q^2=0)=F^*(q^2=0)=1$ explicitly (we suppress the weak
Q-dependence in the nucleon normalization $N_\lambda (Q)$).

\begin{eqnarray}
F_S(q^2)\sim I^2
J_0(q^2) \sim e^{-\frac{1}{2}\frac{a^2a_\lambda^2}{a^2+a_\lambda^2}q^2} \nonumber \\
F_{PS}(q^2)\sim I^2J_1(q^2)/q \sim e^{-\frac{1}{2}\frac{a^2a_\lambda^2}{a^2+a_\lambda^2}q^2}
\nonumber \\
F_S^*(q^2)\sim {I^*}^2J_1^*(q^2)/q \sim e^{-\frac{1}{2}\frac{{a^*}^2a_\lambda^2}{{a^*}^2+a_\lambda^2}q^2} \nonumber \\
F_{PS}^*(q^2)\sim {I^*}^2J_2^*(q^2)=
e^{-\frac{1}{2}\frac{{a^*}^2 a_\lambda^2}{{a^*}^2+a_\lambda^2}q^2}\left(1+\frac{a_\lambda^4}{3({a^*}^2+a_\lambda^2)}q^2\right)
\end{eqnarray}

Similarly, upon collecting all the normalization and re-coupling factors we find the coupling constants to the $N^*$
vertices

\begin{eqnarray}
\frac{f_s^*}{m_s}&&= g_s \, \, (a \, C)\nonumber \\
g_{ps}^*&&= \left(\frac{f}{m}\right)_{PS}\left(\frac{C}{a}\right) \nonumber \\
\frac{f_v^*}{m_v}&&= g_v \, \, (a \, C)\nonumber \\
g_v^*&&=\left(\frac{f}{m}\right)_v\left(\frac{C}{a}\right).
\end{eqnarray}
\noindent
with the universal scaling constant

\begin{equation}
C=\sqrt{\frac{16}{3}}\left(\frac{a \, \, a^*}{a^2+{a^*}^2}\right)^{5/2}.
\end{equation} 

\vspace{1.2cm}

{\bf III. Results and discussion} \\

Characteristic results of our model calculations are summarized in Table 1 and in Figs. (1-5). In Table 1 we list
for $\lambda NN$ coupling constants from \cite{12}, together with the resulting $\lambda NN^*$ constants for the baryon
parameters $a=a^*=0.7fm$. We compare these findings with results from other sources (such as an extraction from the
partial $N^*\rightarrow N\lambda$ decay widths, vector dominance\cite{25} or simple scaling laws; the corresponding
references are given in the legend of table 1; their graphical comparison is given in Fig. 1). Not unexpected, there is no quantitative consistency between the
various models (even the fairly direct determination of the coupling strengths from the corresponding
$N^*$ partial decay widths in first order suffers from large experimental uncertainties\cite{5}). Here a real test
has to be awaited for from a systematic confrontation with experimental data.\\

For the form factors we are guided for the typical size parameters $a \sim a^* 
\sim 0.5 - 1$ fm for the
quark-diquark representation of the baryons from a fit to the proton charge form
factor in the impulse approximation in the range $0<q<1GeV/c$: from a 
comparison with the standard dipole representation with $\Lambda^2=0.71
GeV^2$\cite{25,26}.

\begin{equation}
F_Q(q^2)\approx\left(\frac{\Lambda^2}{\Lambda^2+q^2}\right)^2;
\end{equation}

\noindent we determine qualitatively $a\approx 0.7fm$ (Fig.2) and allow for the $a^*$ an additional variation up to
20\%, expecting a moderate increase in the $N^*$ radius compared to the $N$ from the p-wave excitation. The
resulting form factors - except for the pseudo-scalar $NN^*$ vertex - show a similar typical structure (Figs.(3,4));
the momentum dependence in the fall off reflects only the moderately different baryon size parameters (note that
we still keep the meson as point-like objects). The influence of Lorentz-quenching is significant: it enhances the
ratio $F((1GeV/c)^2)/F(0)$ without quenching by typically a factor of 2; the effective size
parameter varies from its value $a$ for a nucleon at rest to

\begin{equation}
a^2_{eff}=\frac{1}{2}(a^2+a_\lambda^2)=a^2\frac{M^2+q^2/2}{M^2+q^2}=\frac{3}{4}a^2
\end{equation}

\noindent
for $q^2=M^2$ (Fig.(2.b)).\\

A quantitatively different structure is found for the pseudo-scalar and the space-like vector-meson $NN^*$ vertices
(Fig.(5)): while exponentially they exhibit a similar, Gaussian-dominated behavior such as the scalar and time-like
vector form factors, their overall momentum spectrum is enhanced: the overlap of the p-wave in the $N^*$ with the pseudo scalar (p-wave) operator, adds a $q^2$ to the overlap integral of the form factors, which leads to a typical momentum dependence (for $a=a^*=a_{\lambda}$)

\begin{equation}
F_{PS}(q^2)\sim \left(1+\frac{1}{6}a^2q^2\right)e^{-a^2q^2/4};
\end{equation}

\noindent
Of course, for a quantitative insight into the structure of the different form factors a more realistic calculation is necessary.

\vspace{1.2cm}

{\bf IV. Summary and outlook}\\

Summarizing our brief report, we formulated the dominant meson-$NN$ and
meson-$NN^*$ vertices in a simple quark-diquark model for
baryons, together with the assumption of the coupling of mesons as elementary
objects (a philosophy which strongly resembles the
spirit of the quark-coupling-models \cite{27}). We extract both
the relative phases between the $\lambda NN^*$ vertices (for p-wave dominated
$(1/2^-)N^*$ resonances) and the momentum spectra of the hadronic form
factors for momenta up to 1 GeV/c. For our findings we feel on a safe ground for the relative phases of the
coupling constants, however, the hadronic form factors should be improved in the
light of more sophisticated representations for $(1/2^{-})N^*$ resonances (we remark,
however, that in contrast to the merits of our simple two-body representation of
baryons the resolution into 3 quarks immediately raises the well known problems
of center-of-mass corrections and Lorentz-quenching \cite{28}). We expect that with continuous and sophisticated data from hadron and electron factories a
much more profound understanding of the  nature of baryon resonances may be gained in the
near future.\\

In this note we did not address a further serious problem for the relative phases
of the various meson exchange: i.e. the coupling of the $N^*$ vertices to the
meson-nucleon continuum above the pion threshold. We are presently investigating
this aspect - which gives rise to complex form factors - in a simple
perturbative model, hoping for a more consistent understanding of meson-baryon
vertices and thus of the structure $N^*$ resonances high in the continuum of the
nucleon\cite{29}.

\vskip 2.0cm

{\bf Appendix}\\

Here we collect the different overlap integrals for the hadronic form factors.

\begin{eqnarray}
I=\int^\infty_{-\infty}e^{-a^2q^2}dq=\frac{\sqrt{\pi}}{a},
\end{eqnarray}

\begin{eqnarray}
I^*=\int^\infty_{-\infty}e^{-\frac{1}{2}(a^2+{a^*}^2)q^2}dq=\sqrt{\frac{2\pi}{a^2+{a^*}^2}},
\end{eqnarray}

\begin{eqnarray}
J_0(q^2)=\int^\infty_{-\infty}e^{-\frac{1}{2}a^2{q^\prime}^2-\frac{1}{2}a^2_\lambda (q^\prime-q)^2}dq^\prime=\sqrt{\frac{2\pi}{a^2+{a_\lambda}^2}}G(q^2),
\end{eqnarray}

\begin{eqnarray}
J_1(q^2)=\frac{1}{q}\int^\infty_{-\infty}q^\prime e^{-\frac{1}{2}a^2{q^\prime}^2-\frac{1}{2}a^2_\lambda (q^\prime-q)^2}dq^\prime=\sqrt{\frac{2\pi}{a^2+{a_\lambda}^2}}\frac{a_\lambda^2}{a^2+a_\lambda^2}G(q^2),
\end{eqnarray}

\begin{eqnarray}
J_2(q^2)&&=\int^\infty_{-\infty}{q^\prime}^2e^{-\frac{1}{2}a^2{q^\prime}^2-\frac{1}{2}a^2_\lambda
(q^\prime-q)^2}dq^\prime \nonumber \\
&&=\sqrt{\frac{2\pi}{a^2+{a_\lambda}^2}}\frac{1}{a^2+a_\lambda^2}\left(1+\frac{a_\lambda^4}{3(a^2+a_\lambda^2)}q^2\right)G(q^2),
\end{eqnarray}

\noindent
with

\begin{equation}
G(q^2)=e^{-\frac{1}{2}\frac{a^2a_\lambda^2}{a^2+a_\lambda^2}q^2},
\end{equation}

\noindent
(for the $N^*$ vertices $a^2$ has to be replaced by ${a^*}^2$ in $J_i(q^2)$).

\section{Table and Figure Captions}

\underline{Table 1}: Comparison of the $\pi$, $\eta$, $\varrho$, $\omega$,
$\sigma$ and $\delta$ coupling constants
from Gedalin et. al. and Vetter et. al. \cite{8} with the results
from the present investigation (denoted by 'Gedalin', 'Vetter' and 'This 
Calc.)', respectively. The NN coupling constants (denoted by 'NN') are
taken from ref. \cite{12}.

\underline{Figure 1}: Graphical comparison of the various $NN^*-meson$ coupling constants from Table 1. ('This Calc.': boxes, 'Gedalin': circles, 'Vetter': crosses).

\underline{Figure 2}: Extraction of the quark-diquark size parameter $a$ for the
nucleon from a fit of the charge form factor of the proton at momentum transfers
$q<1GeV/c$.

\underline{Figure 3}: Comparison of the scalar $NN$ and $NN^*$ form factors
 for a=0.7 fm and $a^*$=0.6 fm.

\underline{Figure 4} Scalar $NN^*$ form factor for $a$=0.7 fm and $a^*$=0.6/0.8 fm for a the N (and the $N^*$) in rest versus a nucleon with momentum Q = 1 GeV/c.

\underline{Figure 5}: Comparison of the momentum dependence of the pseudo-scalar $NN^*$ and $NN$ form factors, with
$a=0.7$ and $a^*=0.8$, without and including Lorentz quenching.

\begin{table}[ht]
\begin{center}
\begin{tabular}{|c|c|c|c|c|c|cl|}
\hline
 $g_{\lambda NB}$ & $\pi$ & $\eta$  & $\varrho$ & $\omega$ & $\sigma$ &$\delta$& \\
 \hline
 NN & $\sqrt{14.6\times 4\pi}$ & $\sqrt{2.2\times 4\pi}$ & $\sqrt{0.95\times
 4\pi}$ & $\sqrt{20\times 4\pi}$ & $\sqrt{8.03\times 4\pi}$ & $\sqrt{5.07\times 4\pi}$ &\\
This Calc. & 0.908 & 0.81859 & 0.987 & 0.9635 & 0.65755 & 0.9338 & \\
Gedalin & 0.8  & 2.2  & 1.66  & 0.94 & 0.5 & 1.48 & \\
Vetter & 0.79 & 2.22 & 0.615 & 0.236 &     &      & \\
\hline
\end{tabular}
\end{center}
\end{table}

\newpage

\begin{center}
\begin{figure}[htb]
\vspace*{10pt} 
\vspace*{1.4truein}             
\vspace*{10pt} \parbox[h]{4.5cm}{ \includegraphics{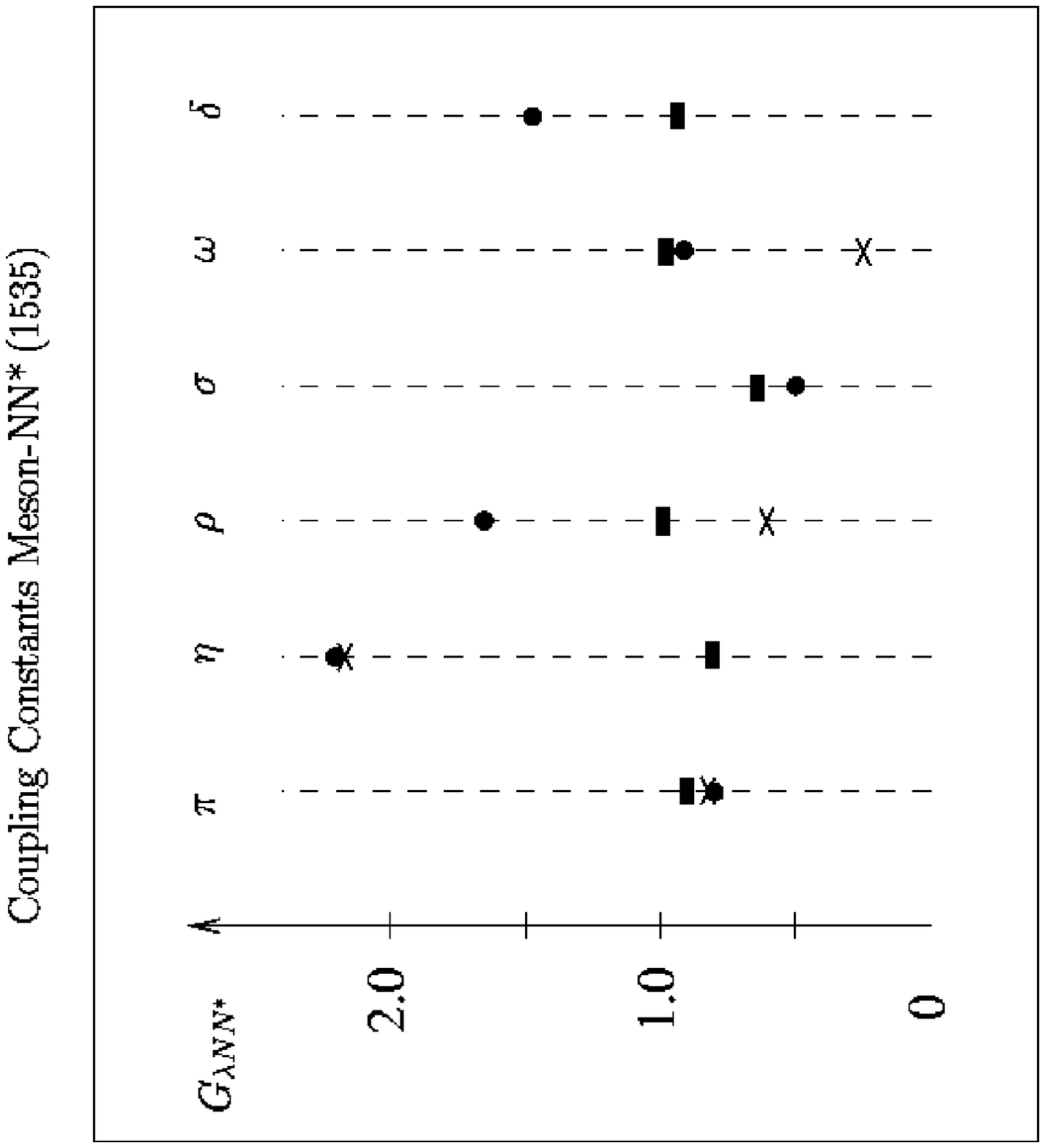}}
\vspace{20pt} \caption{}
\end{figure}
\end{center}

\vspace*{30pt}

\begin{center}
\begin{figure}[htb]
\vspace*{10pt} 
\vspace*{1.4truein}             
\vspace*{10pt} \parbox[h]{4.5cm}{ \includegraphics{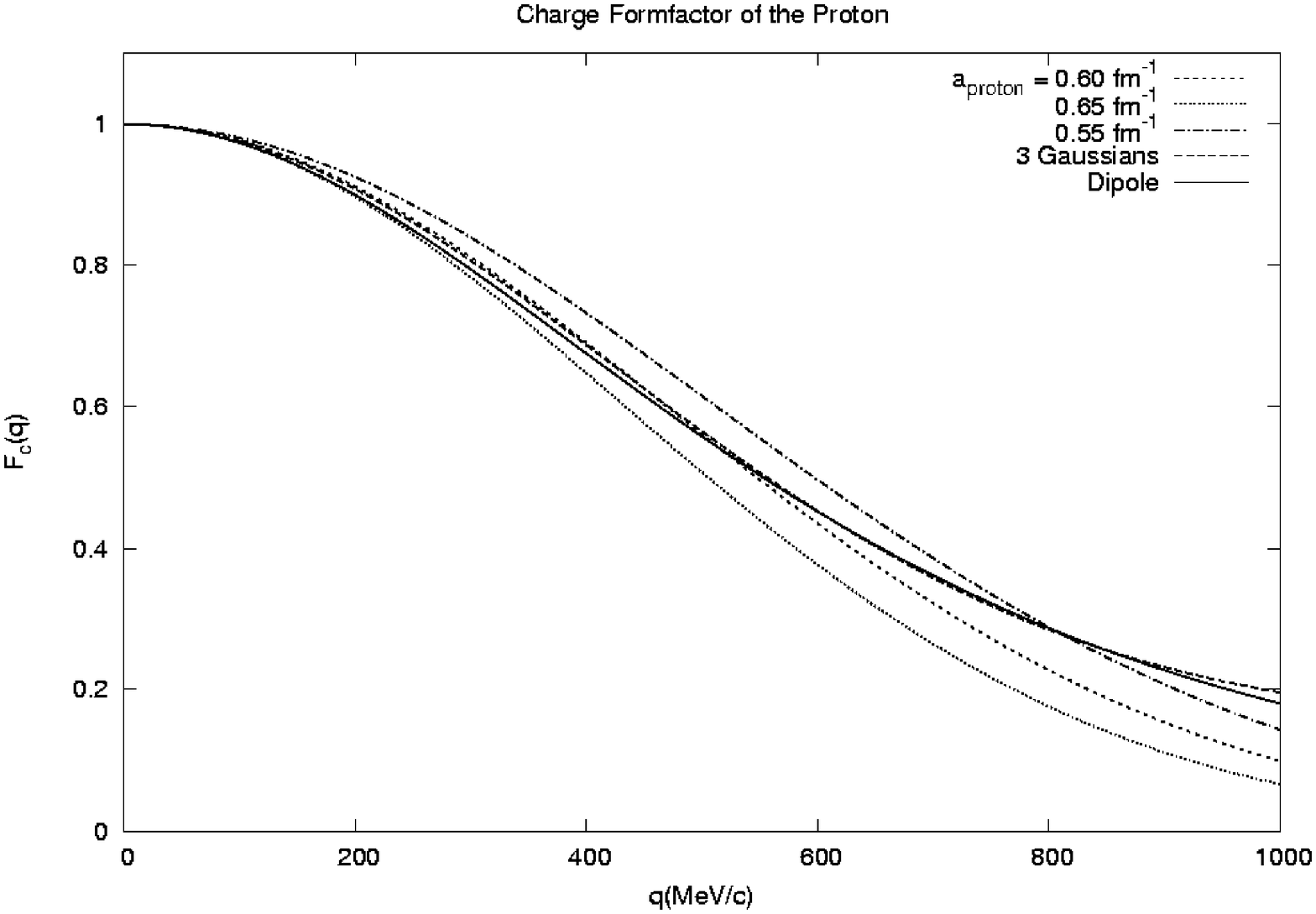}}
\vspace{20pt} \caption{}
\end{figure}
\end{center}

\vspace*{30pt}
\begin{center}
\begin{figure}[htb]
\vspace*{10pt} 
\vspace*{1.4truein}             
\vspace*{10pt} \parbox[h]{4.5cm}{ \includegraphics{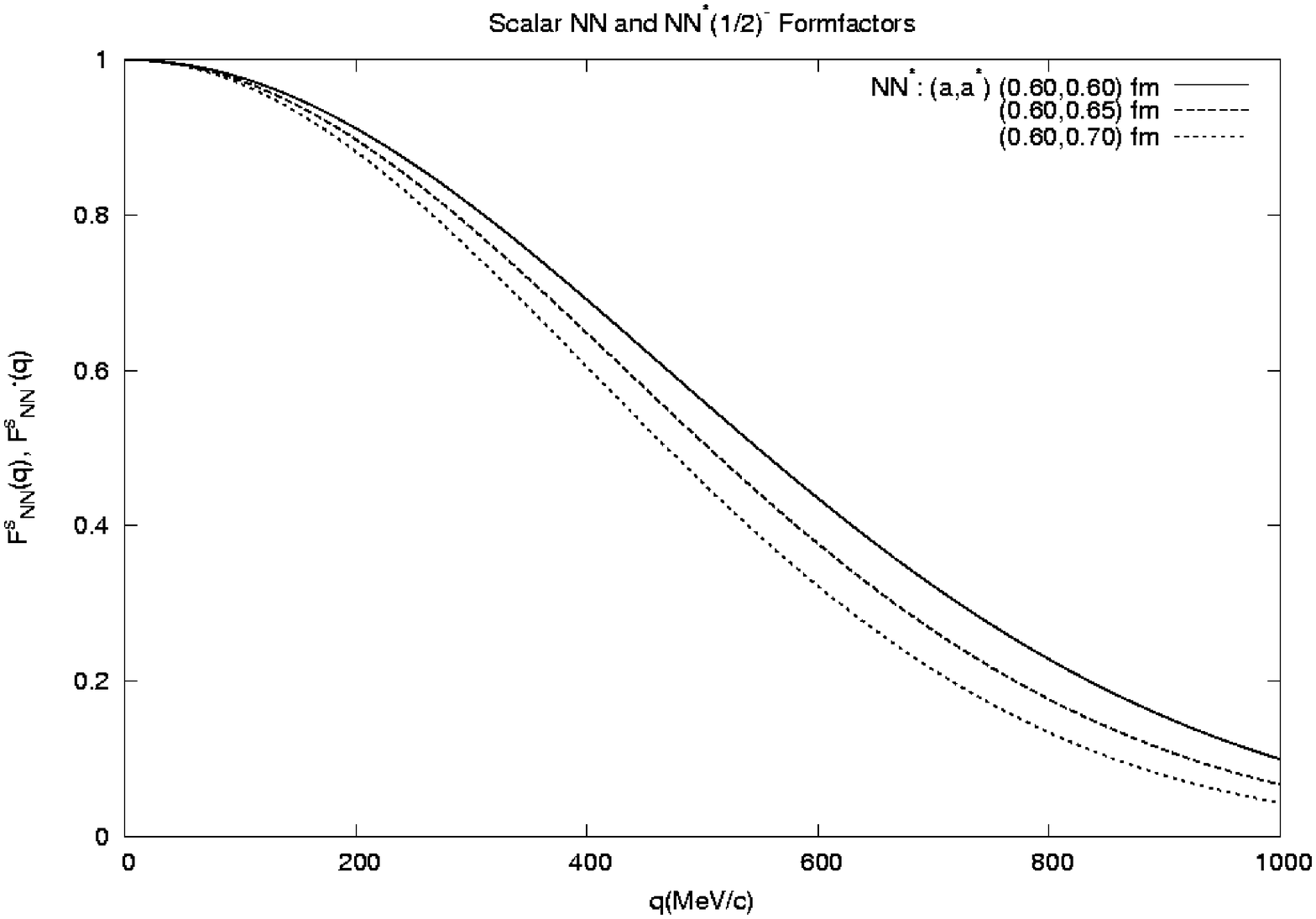}}
\vspace{20pt} \caption{}
\end{figure}
\end{center}

\begin{center}
\begin{figure}[htb]
\vspace*{10pt} 
\vspace*{1.4truein}             
\vspace*{10pt} \parbox[h]{4.5cm}{ \includegraphics{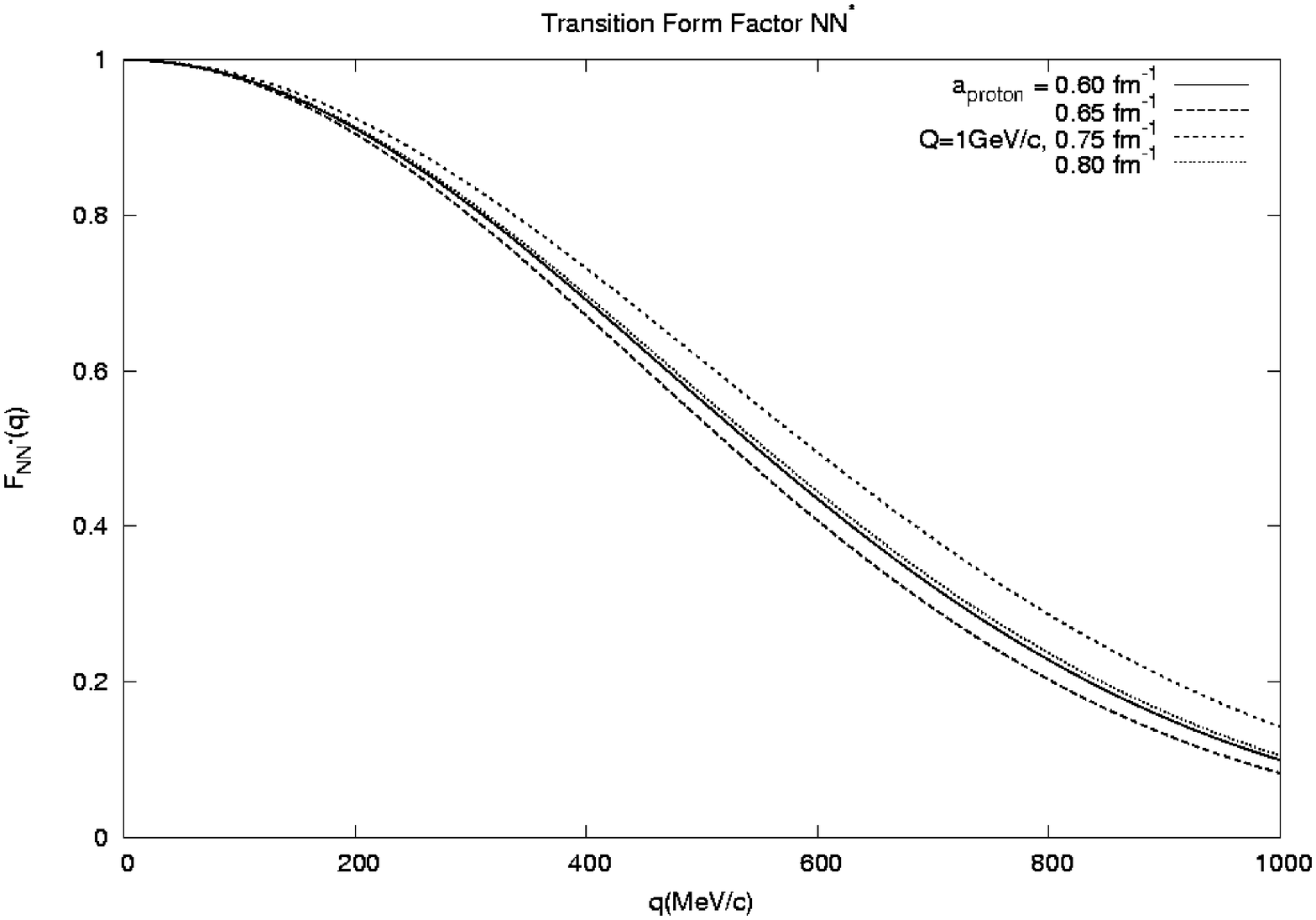}}
\vspace{20pt} \caption{}
\end{figure}
\end{center}

\begin{center}
\begin{figure}[htb]
\vspace*{10pt} 
\vspace*{1.4truein}             
\vspace*{10pt} \parbox[h]{4.5cm}{ \includegraphics{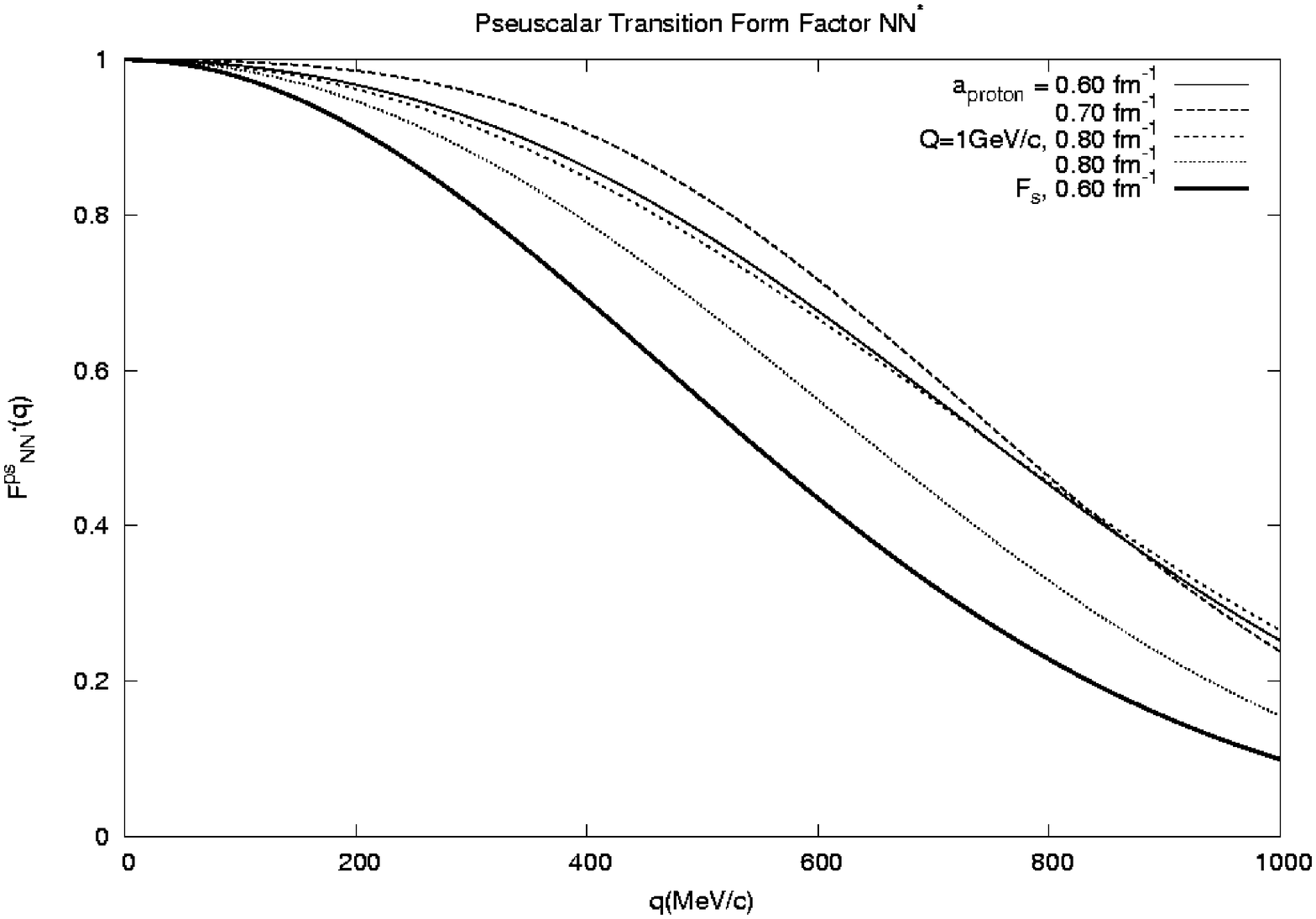}}
\vspace{20pt} \caption{}
\end{figure}
\end{center}

\end{document}